\documentclass[twocolumn]{article}

\usepackage{graphicx}
\usepackage{amsmath,amsfonts}
\usepackage{url}
\usepackage{cite} % For numerical citations
\usepackage{geometry}
\usepackage{setspace}
\usepackage{subfig} % For \subfloat command
\usepackage{balance} % To balance columns on the last page
\usepackage{titlesec}
\usepackage{authblk}

\titleformat{\section}
  {\normalfont\large\bfseries}{\thesection}{1em}{}
\titleformat{\subsection}
  {\normalfont\bfseries}{\thesubsection}{1em}{}
\title{\bfseries The Microwave Rainbow: How Geometry Paints \\ Colours in Microwave Vision}

\author[1*]{Huizhang Yang}
\author[2]{Zhongling Huang}
\author[1]{Zhong Liu}
\author[3]{Jian Yang}

\affil[1]{\small School of Electronic and Optical Engineering, Nanjing University of Science and Technology, Nanjing 210094, China.}
\affil[2]{\small School of Automation, Northwestern Polytechnical University, Xian 710072, China.}
\affil[3]{\small Department of Electronic Engineering, Tsinghua University, Beijing 100084, China.}
\affil[*]{\small e-mail: hzyang@njust.edu.cn}

\date{}

\begin{document}

\twocolumn[
  \begin{@twocolumnfalse}
    \maketitle
\begin{abstract}
Microwave vision from spaceborne synthetic aperture radar (SAR) provides an all-weather, day-and-night capability to observe Earth, yet much of the information encoded in its signals remains undeciphered. Recent high-resolution imagery has revealed a striking phenomenon: man-made structures systematically appear in a spectrum of colours, the physical origin of which has been an open question. Here we show that this effect, which we term the ``microwave rainbow", is a form of geometric dispersion arising from structures acting as intrinsic diffraction gratings. We introduce a geometric-physical model that provides a direct analytical link between a target's geometry and its observed colour signature. This model quantitatively explains the full range of signatures, from continuous colour gradients on curved surfaces (zero-order diffraction) to repeating spectral patterns from periodic structures (high-order diffraction). This work transforms colour from a visual artefact into a precise measure of physical form, enabling the geometry of both critical infrastructure and natural phenomena to be mapped directly from space. Our findings establish the physical basis for a new remote sensing modality—microwave colour vision—and open a new frontier in how we perceive our world.
\end{abstract}
    \vspace{1cm}
  \end{@twocolumnfalse}
]

% --- Main Text ---
\begin{figure*}[t]
  \centering
\includegraphics[width=1.0\linewidth,height=12cm]{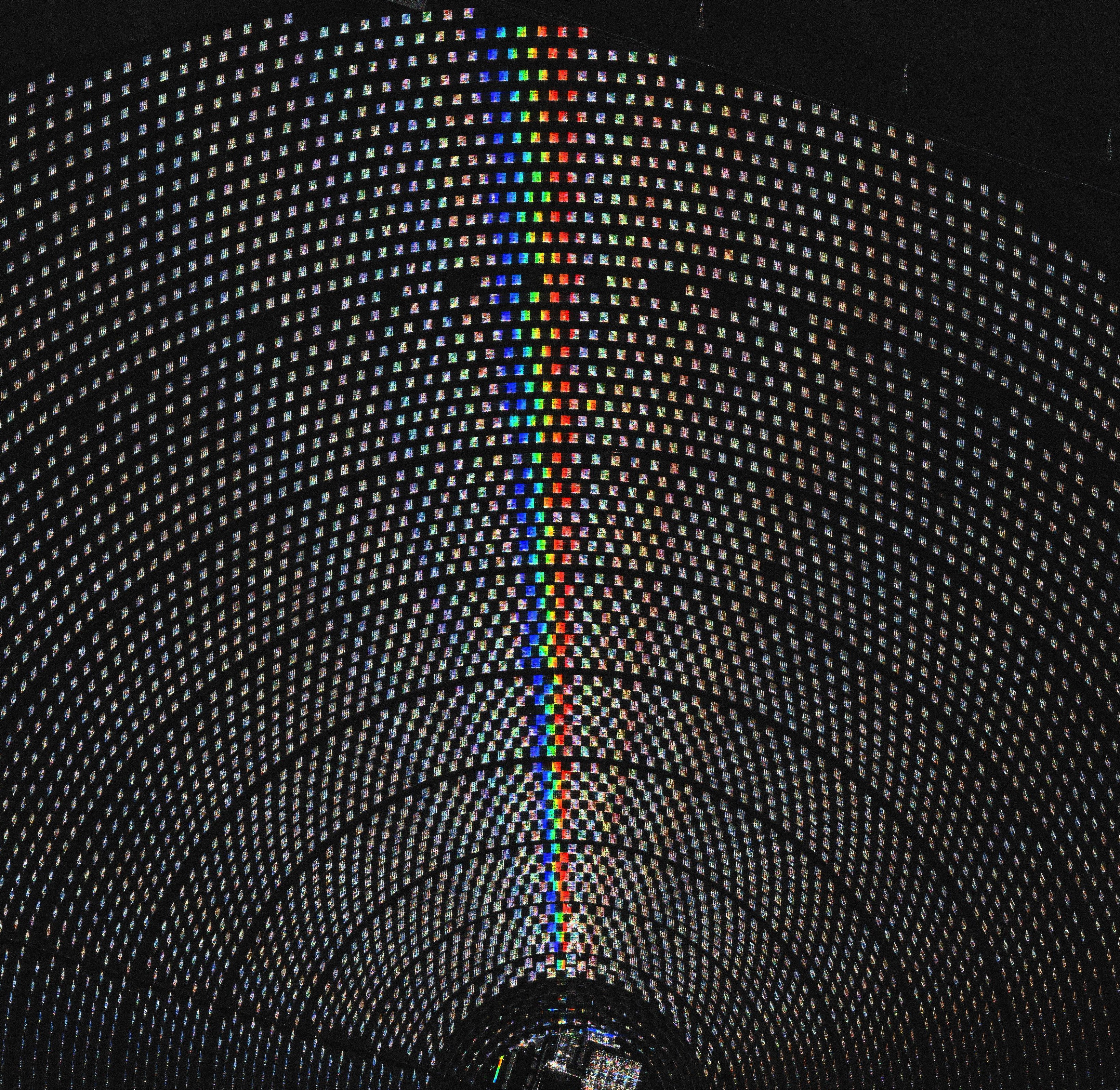}
  \caption{\textbf{A colorized SAR image showing the microwave rainbow.} This effect is analogy to the optical diffraction phenomenon. The imaged area is the concentrating solar power plant, Dunhuang, China, equipped with over 12000 heliostats. The horizontal direction is azimuth, and the vertical direction is range.}
  \label{fig:demo}
\end{figure*}

For decades, we have viewed our planet from space, but only now are we beginning to see it in its true colours---not just in visible light, but in the microwave spectrum. Microwave vision, realised through high-resolution synthetic aperture radar (SAR), grants us the ability to image the Earth's surface at any time of day and in any weather. The advent of very-high-resolution spaceborne systems has pushed this technology to new capabilities, opening new avenues for analysing our built environment \cite{stilla2003potentialAndLimits,gamba2009urbanRemoteSensing,zhu2001vhResolutionSAR, soergel2010radarRemoteSensing, esch2010tanDEMX}. These systems can now resolve infrastructure with decimetre-level detail, making it possible to map cities, monitor construction, and assess structural integrity with high accuracy \cite{ferro2003cosmoSkyMed, corbane2017globalUrbanMapping, dellacqua2012remoteSensingGIS, braun2013terraSARX}.

In these new, incredibly detailed views, a previously unexplained phenomenon has emerged. In novel SAR imaging products, known as colorised sub-aperture images (CSI), the man-made world is rendered in a systematic palette of hues \cite{iceyeNodateCSI, capella2022whatIsCSI, umbra2023csiInterpret}. This effect, which we term the \emph{``microwave rainbow,"} (as shown in Fig. 1), has existed as a systematically unexplained visual artefact. The underlying physical mechanism and model have remained a question, limiting its scientific potential. This raises a question: is this rainbow an aesthetic artefact, or does it encode quantitative information about the structures themselves? This question represents a fundamental gap in our understanding of SAR signal formation, limiting our ability to extract geometric information directly from the colour properties.

In this paper, we address this question by demonstrating that the microwave rainbow is the manifestation of a physical process: geometric dispersion. We establish a formal analogy to optical diffraction, where just as a diffraction grating uses its periodic structure to separate light into its constituent colours via interference, a structure's linear geometry acts as a microwave diffraction grating. It sorts the reflected microwave energy by angle into the Doppler frequency spectrum, which is then visualised as distinct colours.
We have translated this physical insight into a quantitative, predictive model that serves to decode the colour signatures. Our geometric-physic model provides a mathematical link between a target's intrinsic  geometry and its resulting colour signature. It explains both how continuous structures like curved guardrails produce a full rainbow gradient (as a zero-order diffraction effect), and how periodic structures like fences or stadium seating create repeating colour cycles, akin to a high-order diffraction grating.

By establishing the physical basis for these colours, we transform them from a qualitative curiosity into a source of physically interpretable data. Our validated model, proven against a diversity of real-world structures, decodes these observed colorimetric phenomena, allowing us to discern the geometry of critical infrastructure at a planetary scale—from individual buildings and bridges to vast solar power plants and entire city grids. This enables a new form of remote sensing capable of monitoring the geometric fabric of civilization and dynamic natural phenomena like ocean waves. We therefore establish a new remote sensing modality: microwave colour vision, opening a frontier for fine-grained, large-scale monitoring of our planet.

\subsection*{Physical Basis of the Microwave Rainbow}

At the heart of the microwave rainbow lies a principle, which we identify as geometric dispersion. This is the process by which a target's own physical structure acts as an intrinsic filter, sorting reflected energy by angle through coherent interference. The mechanism is analogous to an optical diffraction grating. Just as a grating separates light into a spectrum of colours, a target's  geometry disperses the wideband Doppler signal from the radar, concentrating its energy into a narrow spectral band that is then visualised as distinct colours.

This filtering effect arises from the coherent nature of microwave scattering. A strong, constructive echo is produced only when the radar's line-of-sight is at a specific angle relative to the axis of a linear feature, satisfying the condition for constructive interference. Crucially, in SAR imaging, the viewing angle is intrinsically linked to the Doppler frequency. By sorting the backscattered energy based on angle, the target's geometry is therefore effectively sorting the signal by frequency. A simple geometric line becomes an engine of spectral separation.

This principle explains the colours of the built environment. It governs the response of countless features: from single-bounce specular reflectors like suspended cables and metal guardrails; to the lines of intersection in double-bounce dihedral structures (e.g., a wall-ground junction); and even to periodic arrays of discrete scatterers, such as fence posts or stadium seats, which act collectively as a single, highly directional target. To transform this physical insight into a predictive tool, a quantitative mathematical model is required.

\subsection*{Geometric-Physic Model of the Microwave Rainbow}
To mathematically decode the microwave rainbow, we developed a quantitative framework that links a target's geometry to its unique colour signature. The model is rooted in the fundamental physics of SAR: the sensor's viewing angle, known as the squint angle ($\theta_{sq}$), directly maps to the Doppler frequency ($f_d$) of the radar echo, given by $f_d = (2V/\lambda)\sin(\theta_{sq})$. CSI products visualise this relationship by assigning different portions of the Doppler frequency spectrum to a set of basis colours in RGB hues. Our model provides the quantitative rules that govern this process, explaining how a target's geometric orientation acts as a diffraction grating. This grating effect gives rise to distinct diffraction orders, which direct reflected energy into specific parts of the Doppler spectrum, thus giving the target its colour.

Our model reveals how these diffraction orders manifest in SAR imagery, corresponding to the two distinct types of ``rainbows" observed:
 1) \emph{The Zero-Order Diffraction ($m=0$): Single Rainbows from Continuous Targets.} For a continuous linear target, such as a bridge cable or a curved guardrail, the scattering is dominated by the zero-order diffraction response. This produces a single, well-defined peak in the Doppler spectrum. The radar squint angle ($ \theta_{sq}$) of this peak is governed by the target's effective azimuth orientation angle $\theta_{az}$ in the SAR imaging plane through the core relationship: $\theta_{sq} = -\theta_{az}$. This direct mapping means that as a target's orientation changes smoothly along a curve, its colour signature sweeps seamlessly across the entire spectrum, creating a continuous microwave rainbow.
 2) \emph{The High-Order Diffraction ($m \neq 0$): Repeating Rainbows from Discrete Targets.} In contrast, for discrete periodic targets like fences or ribbed roofing with a physical spacing $d_x$, the coherent interference produces a series of prominent high-order diffraction responses in addition to the zero-order one. This results in multiple, simultaneous peak responses in the angle and Doppler domains. The squint angle for the $m$-th order response is given by the analytical solution:
    \begin{equation}
        \theta_{sq,m} = \arcsin\left( \frac{m\lambda \cos(\theta_{az})}{2d_x} \right) - \theta_{az}
    \end{equation}
where $m$ is the integer diffraction order. The non-zero orders ($m = \pm1, \pm2, \dots$) are responsible for the unique repeating rainbow effect, revealing how a single geometric orientation can appear as multiple distinct colours. As $\theta_{az}$ varies, its signature in a CSI exhibits a periodic colour change, as seen in Fig. \ref{fig:demo}(a).

To apply this model to the real world, we must project a target's 3D geometric orientation onto the 2D SAR imaging plane. By modelling a 3D linear target using its horizontal ($\theta_h$) and vertical ($\theta_v$) orientation angles, we derive its effective 2D orientation ($\theta_{az}$) for a given broadside incidence angle $\theta_{inc}$. This culminates in the equation for predicting the peak of the dominant zero-order response:
\begin{equation}
    \tan(\theta_{sq}) = - \cos(\theta_{\mathrm{inc}}) \left( \tan(\theta_{\mathrm{inc}})\tan\theta_{\mathrm{h}} + \tan\theta_{\mathrm{v}} \right)
\end{equation}
This closed-form expression provides a predictive tool for calculating the exact colour signature (via the peak squint angle) for a linear target. A particularly important case is when this zero-order response occurs at zero squint ($\theta_{sq}=0$), which happens when $\tan(\theta_{\mathrm{inc}})\tan\theta_{\mathrm{h}} + \tan\theta_{\mathrm{v}} = 0$, resulting in a green hue.

We validated our geometric-physic model through a series of numerical simulations. These simulations successfully reproduced the key phenomena of both continuous (zero-order) and repeating (high-order) microwave rainbows, confirming the model's predictive power. Full details of the simulation experiments and their results are presented in the Supplementary Information (Supplementary Figs. 1-2).

\subsection*{Decoding the Microwave Rainbow: From Geometry to Colour}
\begin{figure}[t]
    \centering
    \includegraphics[width=1\columnwidth]{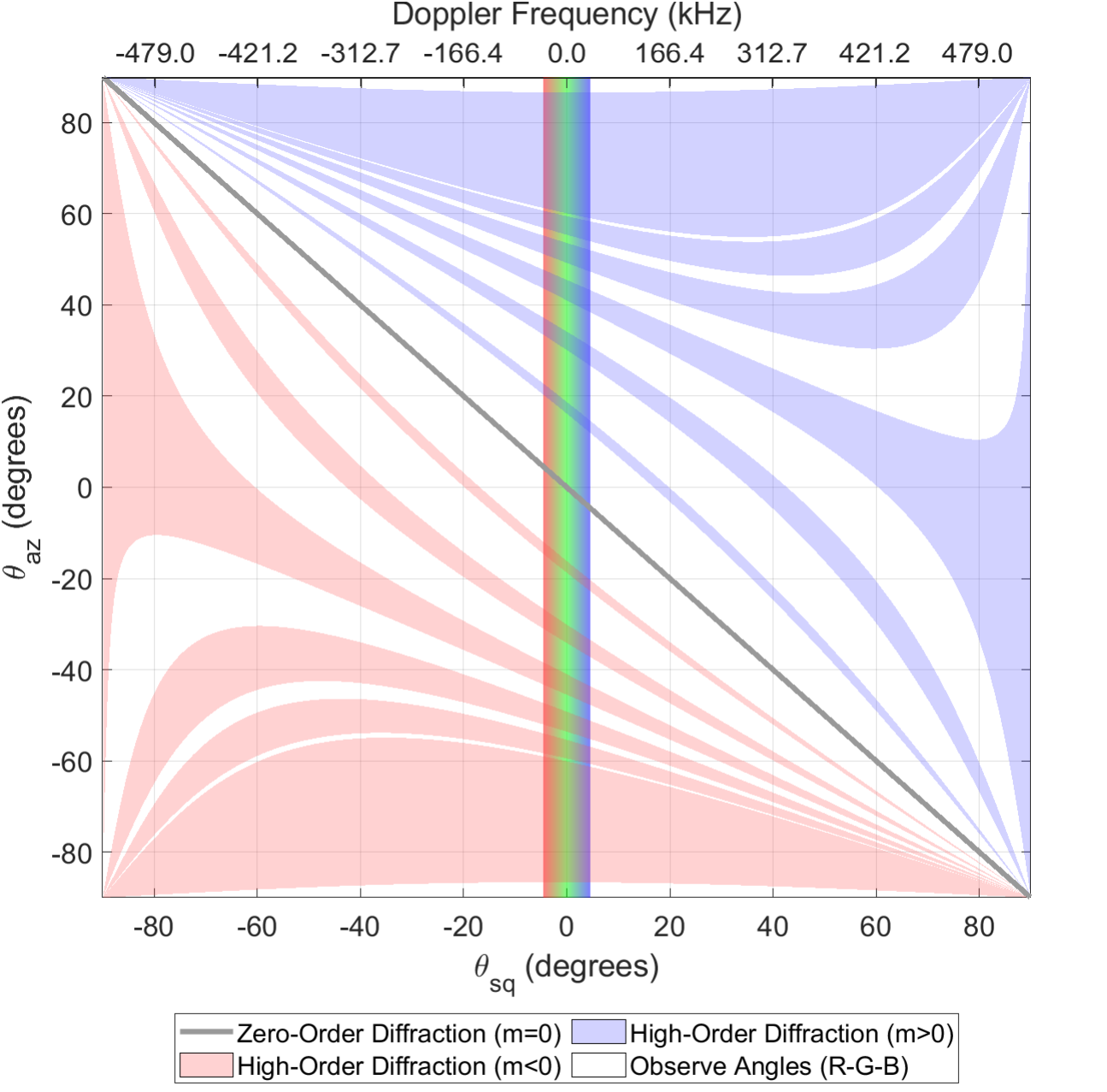}
    \caption{\textbf{Graphical interpretation model for the microwave rainbow.} The plot shows the required target orientation $\theta_{az}$ (y-axis) to produce a peak response at a given observation squint angle $\theta_{sq}$ (bottom x-axis) or Doppler frequency (top x-axis). The grey line represents the main response ($m=0$) for continuous targets, defining the primary colour gradient. The shaded regions represent grating responses ($m\neq0$) for discrete periodic targets, leading to repeating colour patterns. The red-green-blue gradient background visualises the observable Doppler window and its mapping to CSI colour, illustrating the spectral components of the microwave rainbow. Calculation parameters: $f_c=9.6$ GHz, $V=7600$ m/s, spatial  resolution = 0.1 m , $d_x=0.05$ m.}
    \label{fig:dispersion_model_plot}
\end{figure}
% FIGURE 1: SLICED COLOURS
\begin{figure*}[]
    \centering
    % Row 1
    \subfloat[Discrete colours in Las Vegas]{\includegraphics[width=0.24\textwidth,height=4cm]{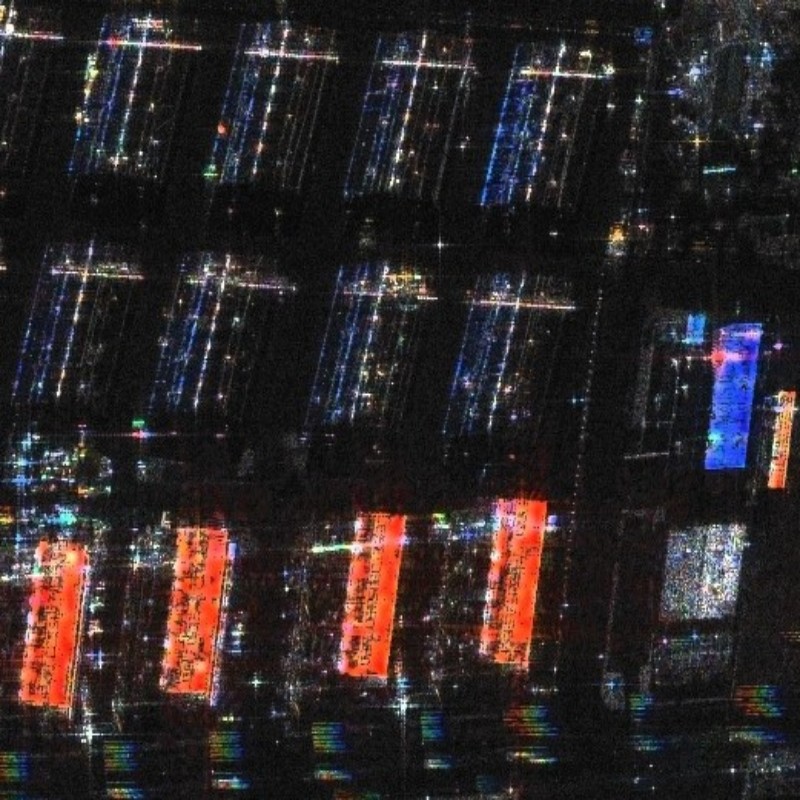}\label{fig:vegas_csi_new}
    \includegraphics[width=0.24\textwidth,height=4cm]{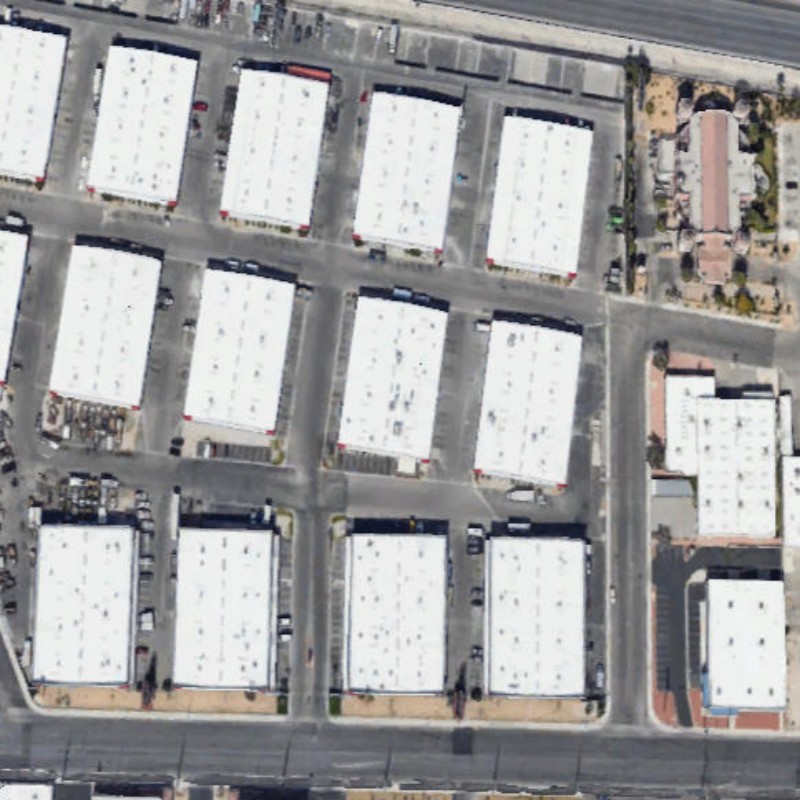}\label{fig:vegas_optical_new}}
    \hfill
    \subfloat[Ribbed roofs]{\includegraphics[width=0.24\textwidth,height=4cm]{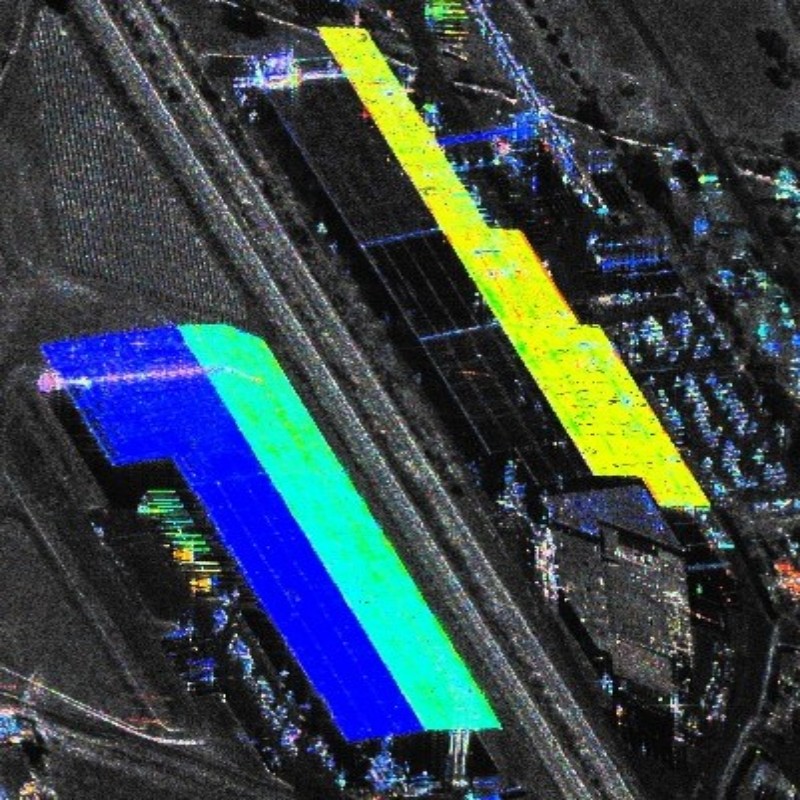}\label{fig:roofs_csi_new}
    \includegraphics[width=0.24\textwidth,height=4cm]{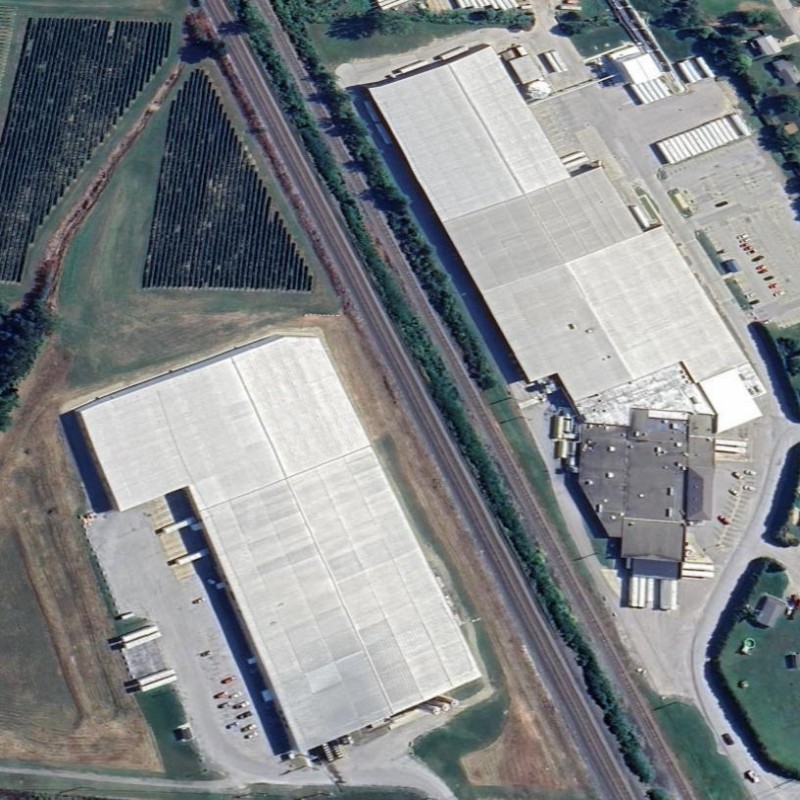}\label{fig:roofs_optical_new}}
    \\ \vspace{1ex}
    % Row 2
    \subfloat[Park MGM, Las Vegas]{\includegraphics[width=0.24\textwidth,height=4cm]{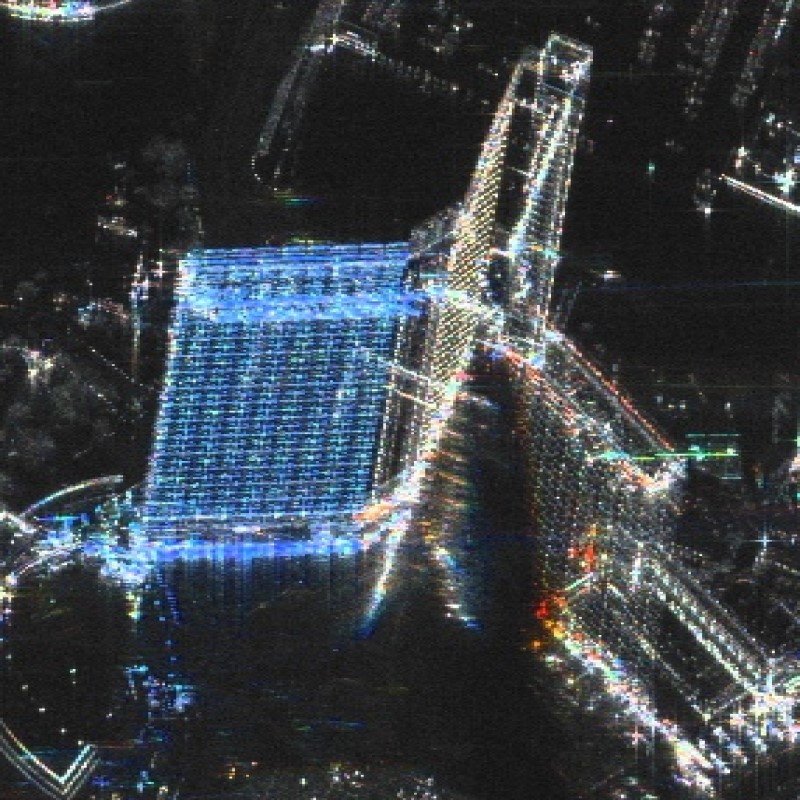}\label{fig:mgm_csi_new}
    \includegraphics[width=0.24\textwidth,height=4cm]{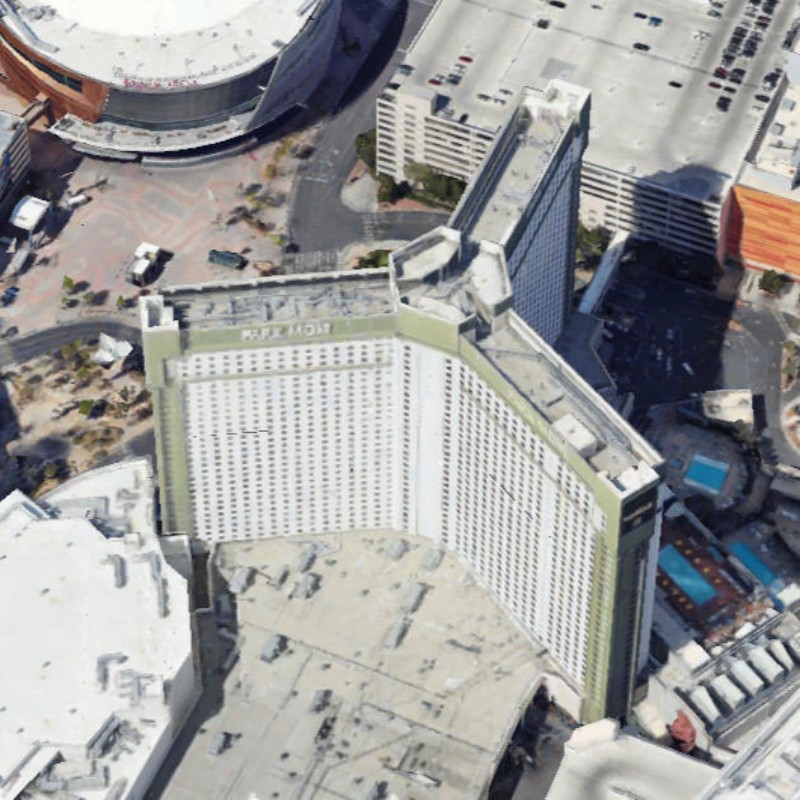}\label{fig:mgm_optical_new}}
    \hfill
    \subfloat[Qiansimen bridge, China]{\includegraphics[width=0.24\textwidth,height=4cm]{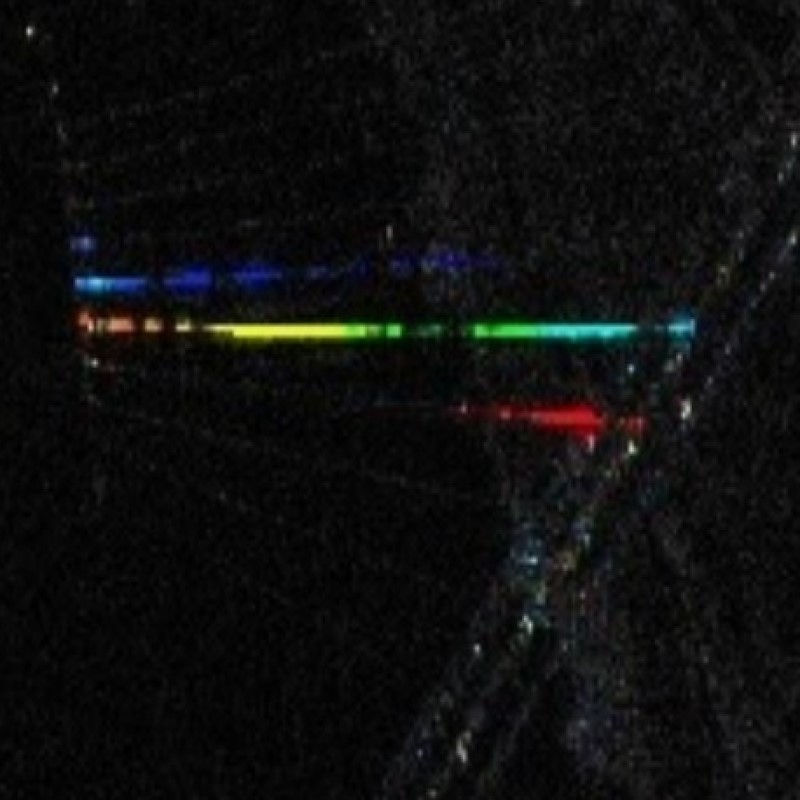}\label{fig:qiansimen_csi_new}
    \includegraphics[width=0.24\textwidth,height=4cm]{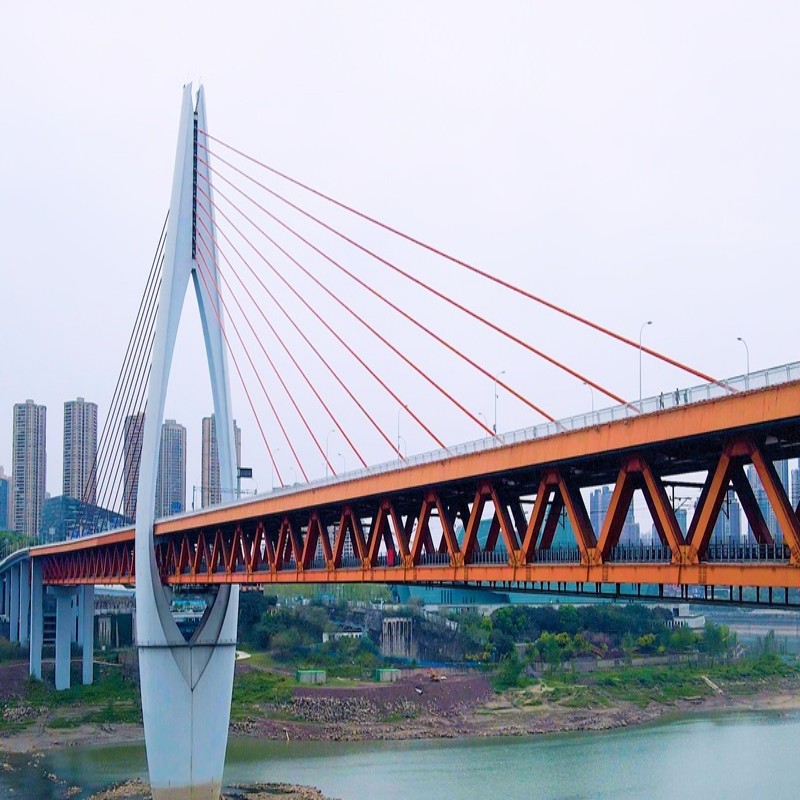}\label{fig:qiansimen_optical_new}}
    
    \caption{\textbf{Sliced colours from the microwave rainbow in man-made structures.} For each pair, the CSI is on the left and the optical view is on the right. \textbf{a}, Different roof orientations on Las Vegas buildings produce distinct red and blue hues. \textbf{b}, Adjacent buildings with differently oriented ribbed roofs show distinct color signatures. \textbf{c}, The specific orientation of the Park MGM facade results in a blue signature. \textbf{d}, Slightly curved steel cables of the Qiansimen Bridge create partial RGB gradients.}
    \label{fig:sliced_colours}
\end{figure*}

% FIGURE 2: CONTINUOUS RAINBOW
\begin{figure*}[]
    \centering
    % Row 1
    \subfloat[Curved highway guardrail, USA]{\includegraphics[width=0.24\textwidth,height=4cm]{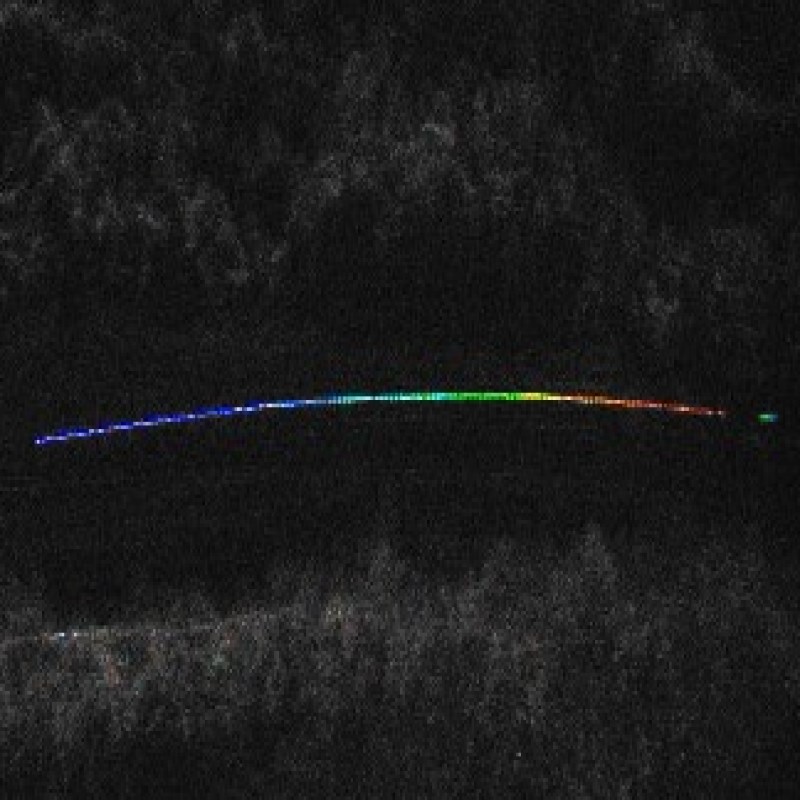}\label{fig:guardrail_csi_new}
    \includegraphics[width=0.24\textwidth,height=4cm]{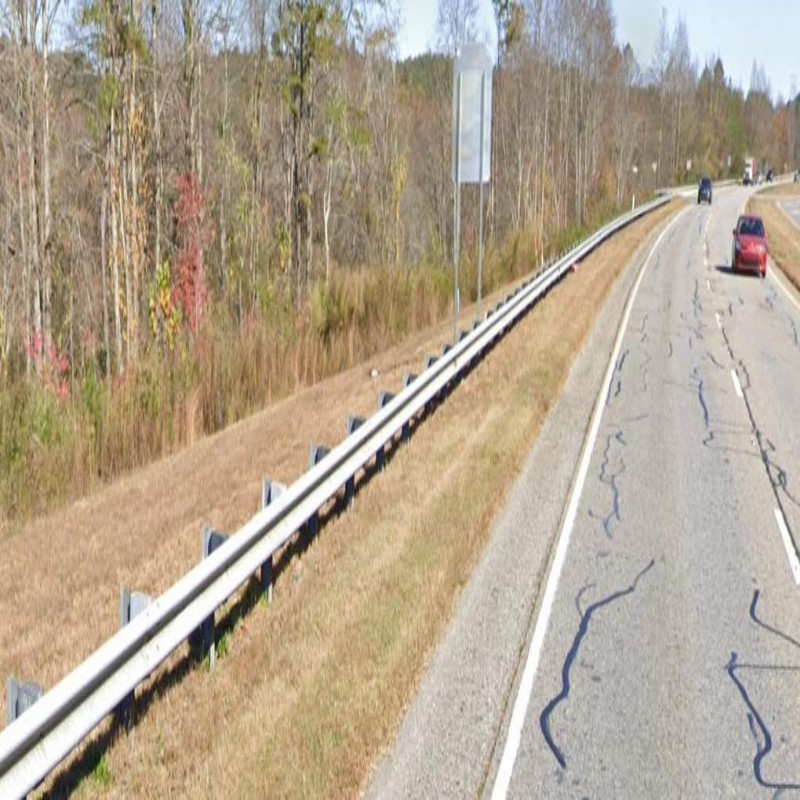}\label{fig:guardrail_optical_new}}
    \hfill
    \subfloat[Catenary power lines, USA]{\includegraphics[width=0.24\textwidth,height=4cm]{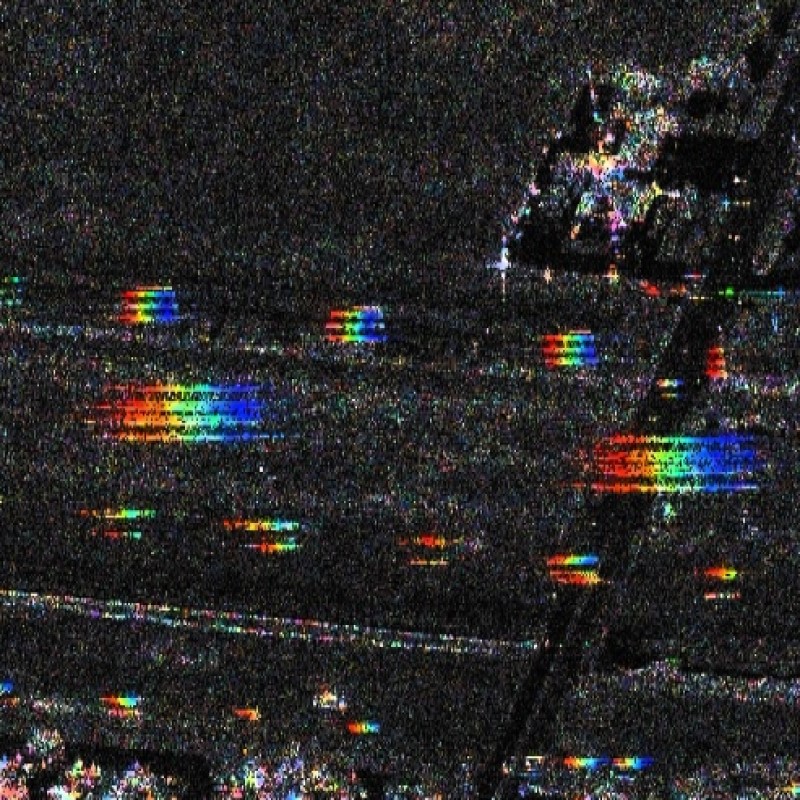}\label{fig:power_lines_csi_new}
    \includegraphics[width=0.24\textwidth,height=4cm]{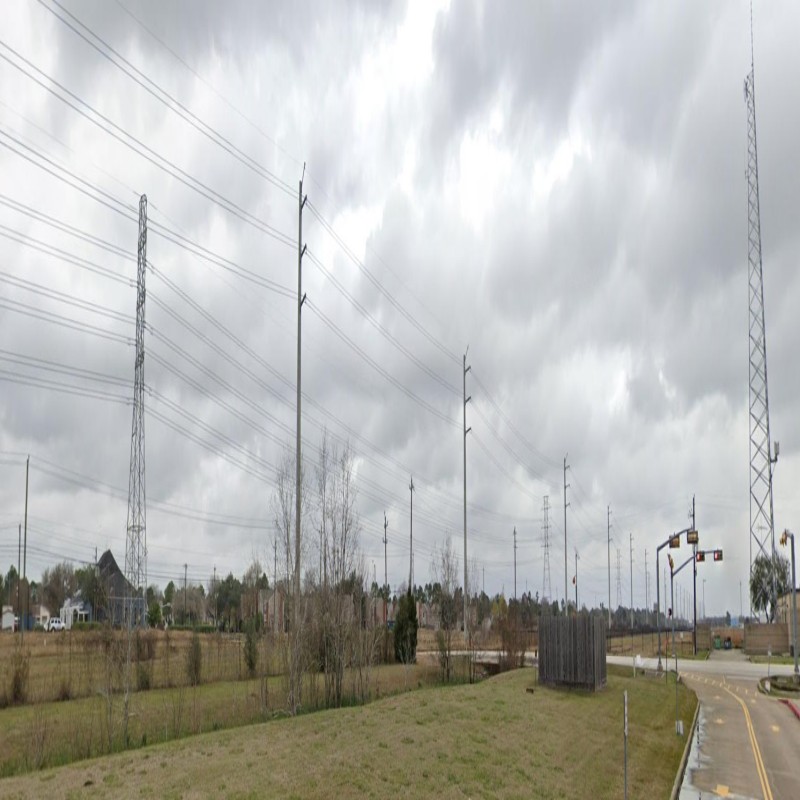}\label{fig:power_lines_optical_new}}
    \\ \vspace{1ex}
    % Row 2
    \subfloat[Dunhuang concentrating solar plant, China]{\includegraphics[width=0.24\textwidth,height=4cm]{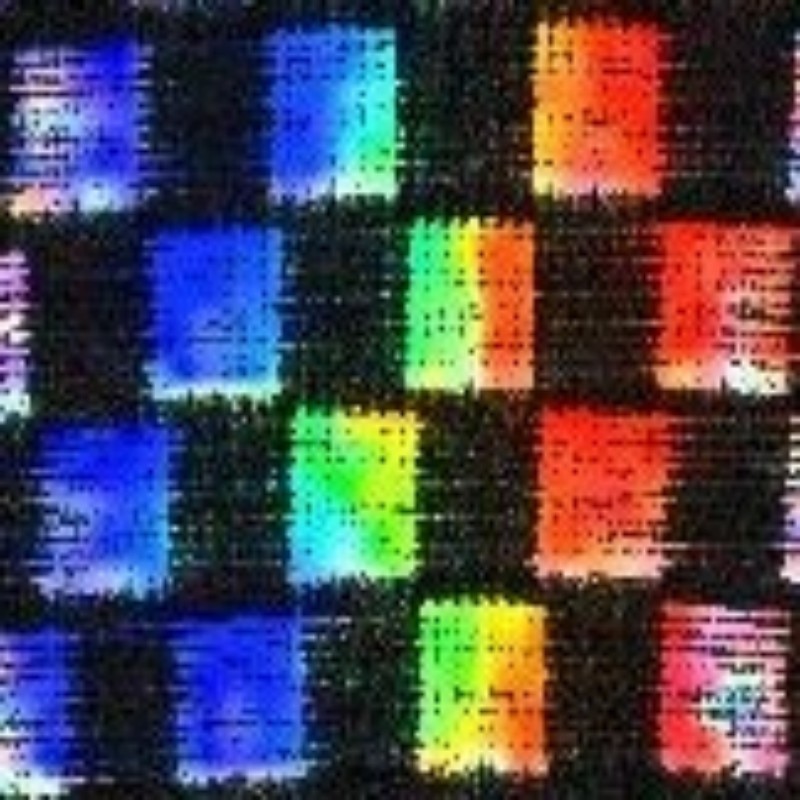}\label{fig:dunhuang_csi_new}
    \includegraphics[width=0.24\textwidth,height=4cm]{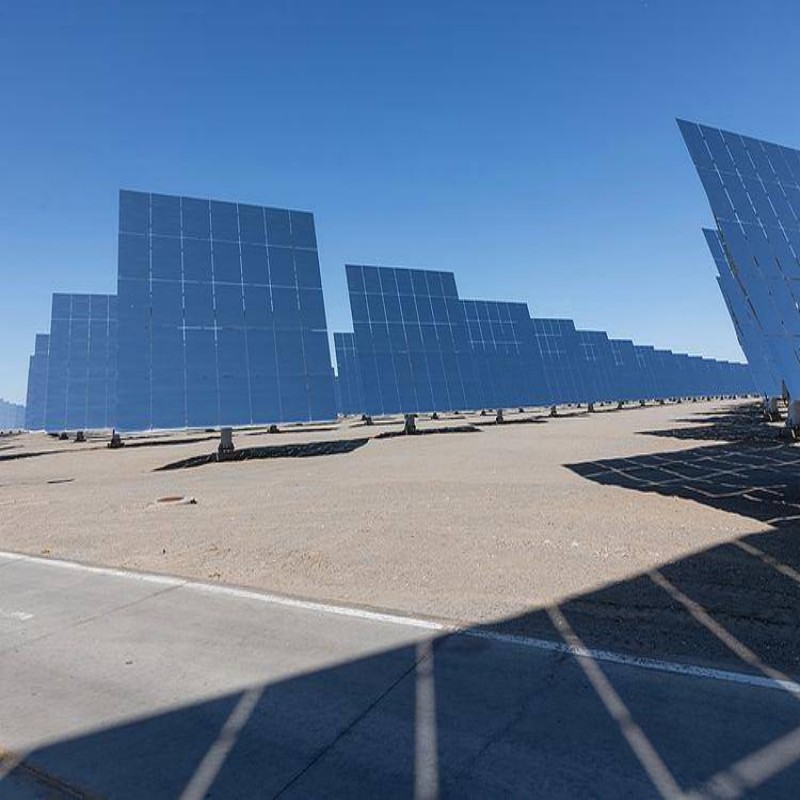}\label{fig:dunhuang_optical_new}}
    \hfill
    \subfloat[Yachi bridge's steel cables, China]{\includegraphics[width=0.24\textwidth,height=4cm]{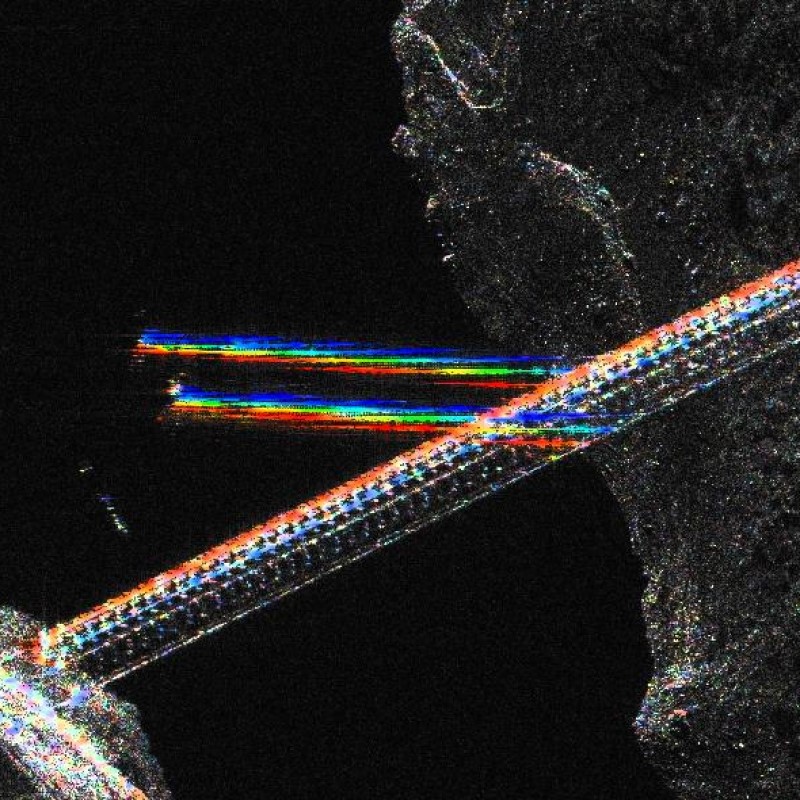}\label{fig:yachi_csi_new}
    \includegraphics[width=0.24\textwidth,height=4cm]{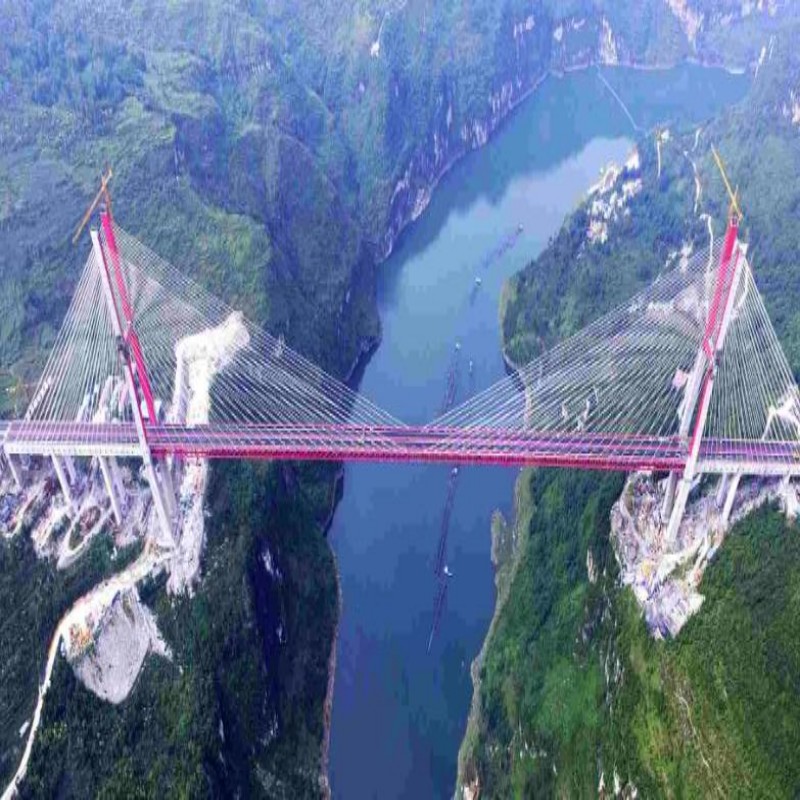}\label{fig:yachi_optical_new}}
    
    \caption{\textbf{Zero-order diffraction microwave rainbows in curved infrastructures.} For each pair, the CSI is on the left and the optical view is on the right. \textbf{a}, A curved steel guardrail produces a full RGB microwave rainbow. \textbf{b}, The catenary sag of power lines creates a continuous spectral rainbow. \textbf{c}, Systematic orientation changes in the heliostat field of a solar power plant generate a full rainbow. \textbf{d}, The stay-cables of the Yachi Bridge show colour gradients due to their gentle curve.}
    \label{fig:continuous_rainbow}
\end{figure*}

% FIGURE 3: REPEATING AND OTHER RAINBOWS
\begin{figure*}[]
    \centering
    % Row 1
    \subfloat[Waterfront fence, Dubai]{\includegraphics[width=0.24\textwidth,height=4cm]{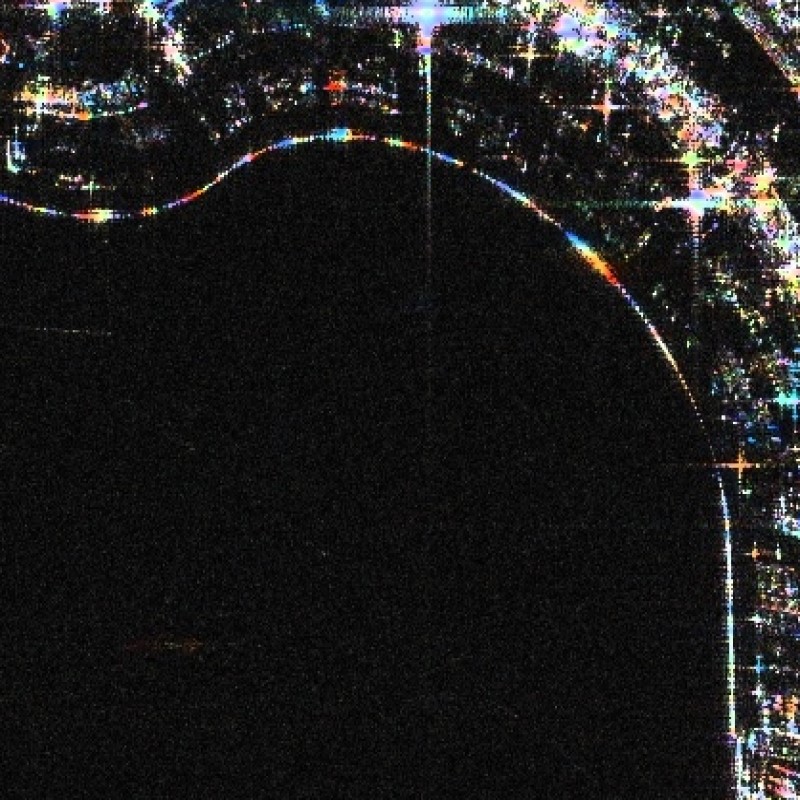}\label{fig:dubai_csi_new}
    \includegraphics[width=0.24\textwidth,height=4cm]{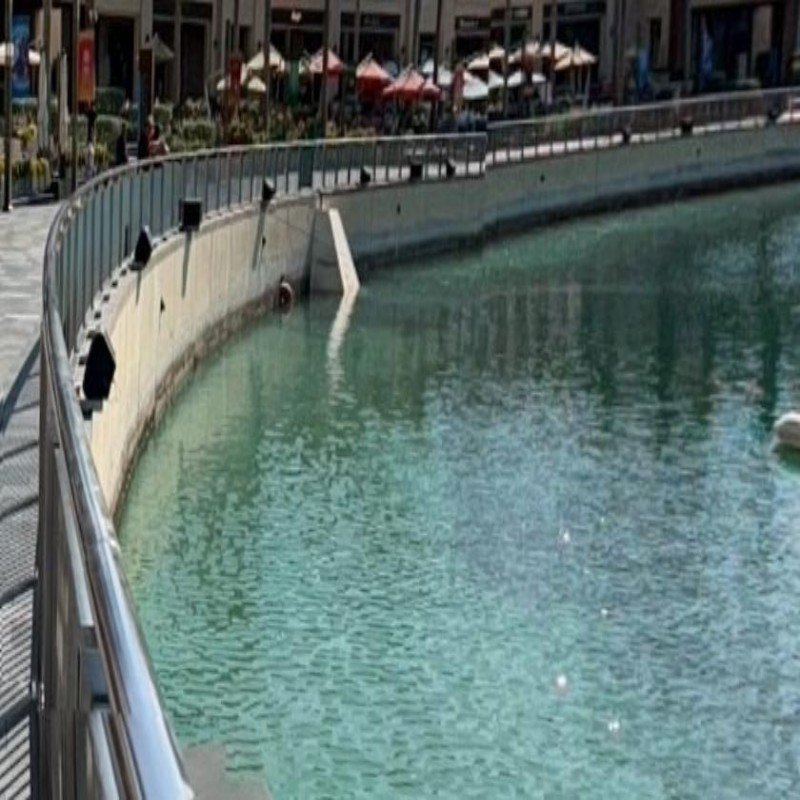}\label{fig:dubai_optical_new}}
    \hfill
    \subfloat[Olympiastadion, Germany]{\includegraphics[width=0.24\textwidth,height=4cm]{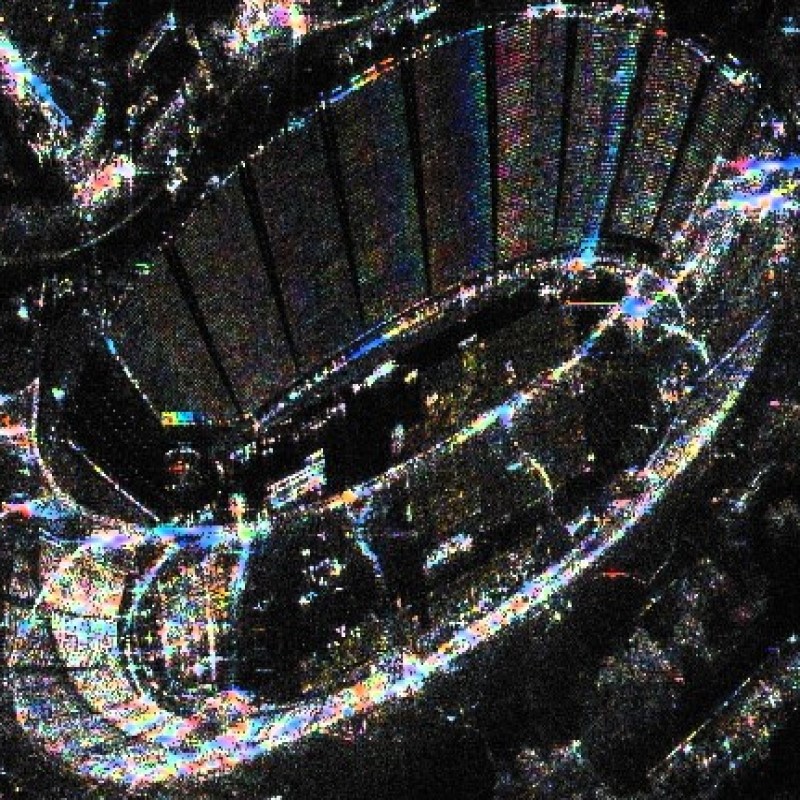}\label{fig:olympiastadion_csi_new}
    \includegraphics[width=0.24\textwidth,height=4cm]{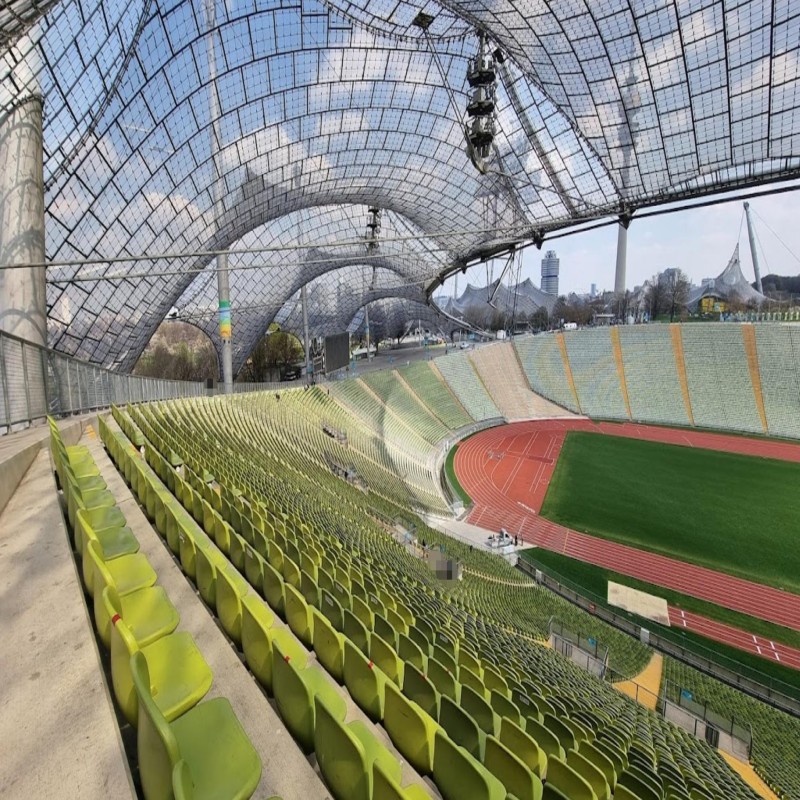}\label{fig:olympiastadion_optical_new}}
    \caption{\textbf{High-order microwave rainbow phenomena.} For each pair, the CSI is on the left and the optical view is on the right. \textbf{a}, A periodic metal fence along a curved waterfront in Dubai produces a repeating RGB→RGB rainbow. \textbf{b}, Curved seating arrays in the Munich Olympiastadion act as a large-scale diffraction grating.}
    \label{fig:repeating_rainbows}
\end{figure*}

Our geometric-physic model allows for the interpretation of the colours observed in SAR imagery, effectively allowing us to ``decode the microwave rainbow." As established by the model, a target's geometry directly dictates its Doppler frequency. CSI products then translate this Doppler frequency into a visible hue by mapping different spectral bands to basis colours. This direct chain of causality—from target geometry to Doppler frequency to observable colour—allows us to interpret the hue of man-made structures as a direct indicator of their orientation. The model predicts the following colour signatures:
 1)  \emph{Green Hue: The Broadside Rainbow.} A target oriented parallel to the ground range axis ($\theta_{az} = 0^\circ$) will have its peak response at zero squint ($\theta_{sq} = 0^\circ$). Its energy is centred around zero Doppler ($f_d= 0$), concentrating its response in the green channel. This results in a green appearance, representing the ``centre" or ``broadside" colour of the rainbow.
2) \emph{Yellow-to-Red Hue: The Rainbow's Red Edge.} A target with a positive orientation angle ($\theta_{az} > 0^\circ$) produces a peak response at a negative squint angle ($\theta_{sq} \le 0^\circ$). This corresponds to a negative peak Doppler frequency ($f_d< 0$), causing the target to appear yellow/red in the CSI. This represents one ``edge" of the rainbow spectrum.
3) \emph{Cyan-to-Blue Hue: The Rainbow's Blue Edge.} Conversely, a target with a negative orientation angle ($\theta_{az} < 0^\circ$) produces its peak response at a positive squint angle ($\theta_{sq} \ge 0^\circ$). This results in a positive peak Doppler frequency ($f_d> 0$), making it appear cyan/blue. This represents the other ``edge" of the rainbow spectrum.
This understanding transforms colour from a mere visual artefact into a quantitative measure of a target's geometric properties.

The behaviour of geometric dispersion, and how it paints the microwave rainbow, can be understood through a graphical visualisation. The plot in Fig.~\ref{fig:dispersion_model_plot} serves as an interpretation tool for our framework. It characterises the precise relationships between target geometry, viewing angle, and Doppler frequency. This diagram addresses the question: \textit{for a given SAR acquisition squint range, what geometric orientations will be ``illuminated" with specific colours in the microwave rainbow, and why?}

The plot is constructed based on the full response equation derived in Eq. (1). The key components of the figure are: 
1) \emph{Dual Axes:} The bottom horizontal axis represents the geometric squint angle ($\theta_{sq}$) in degrees, while the top horizontal axis shows the corresponding physical Doppler frequency in kHz.
2) \emph{Observable Window:} The background with the red-green-blue gradient visually represents the Doppler frequency range captured by the SAR system's azimuth bandwidth ($B_a$). This gradient directly mimics the CSI mapping, illustrating how different spectral parts of the microwave rainbow are assigned specific hues.
3) \emph{Zero-Order Diffraction Region ($m=0$):} The solid grey line represents the zero-order diffraction response, which governs continuous targets. This region dictates the primary colour sweep of the microwave rainbow. For any point on this line, the corresponding $\theta_{sq}$ and $\theta_{az}$ satisfy the core condition $\theta_{sq}= -\theta_{az}$.
4) \emph{High-Order Diffraction Region ($m \neq 0$):} The shaded red and blue regions show the solutions for discrete periodic targets, representing the high-order diffraction responses. They are regions, not lines, because the radar's range bandwidth ($B_r$) broadens the response. These regions demonstrate how a single geometric orientation can produce strong returns at multiple, predictable Doppler frequencies simultaneously, leading to the repeating colour cycles that are a key characteristic of the microwave rainbow.

This graphical model serves as an interpretation tool for the complex colourimetric signatures in SAR imagery. For example, to understand what a discrete target with an orientation of $\theta_{az}=20^\circ$ will look like, one can draw a horizontal line at this value. The intersections of this line with the zero-order path and the high-order response areas reveal all the squint angles (and thus Doppler frequencies/colours) at which this target will produce a strong response, effectively predicting its contribution to the microwave rainbow. This predictive capability is used for decoding the observed colours of complex man-made structures.

\begin{figure*}[]
    \centering
    \subfloat[CSI of Los Angeles urban grid]{\includegraphics[width=0.48\linewidth,height=10cm]{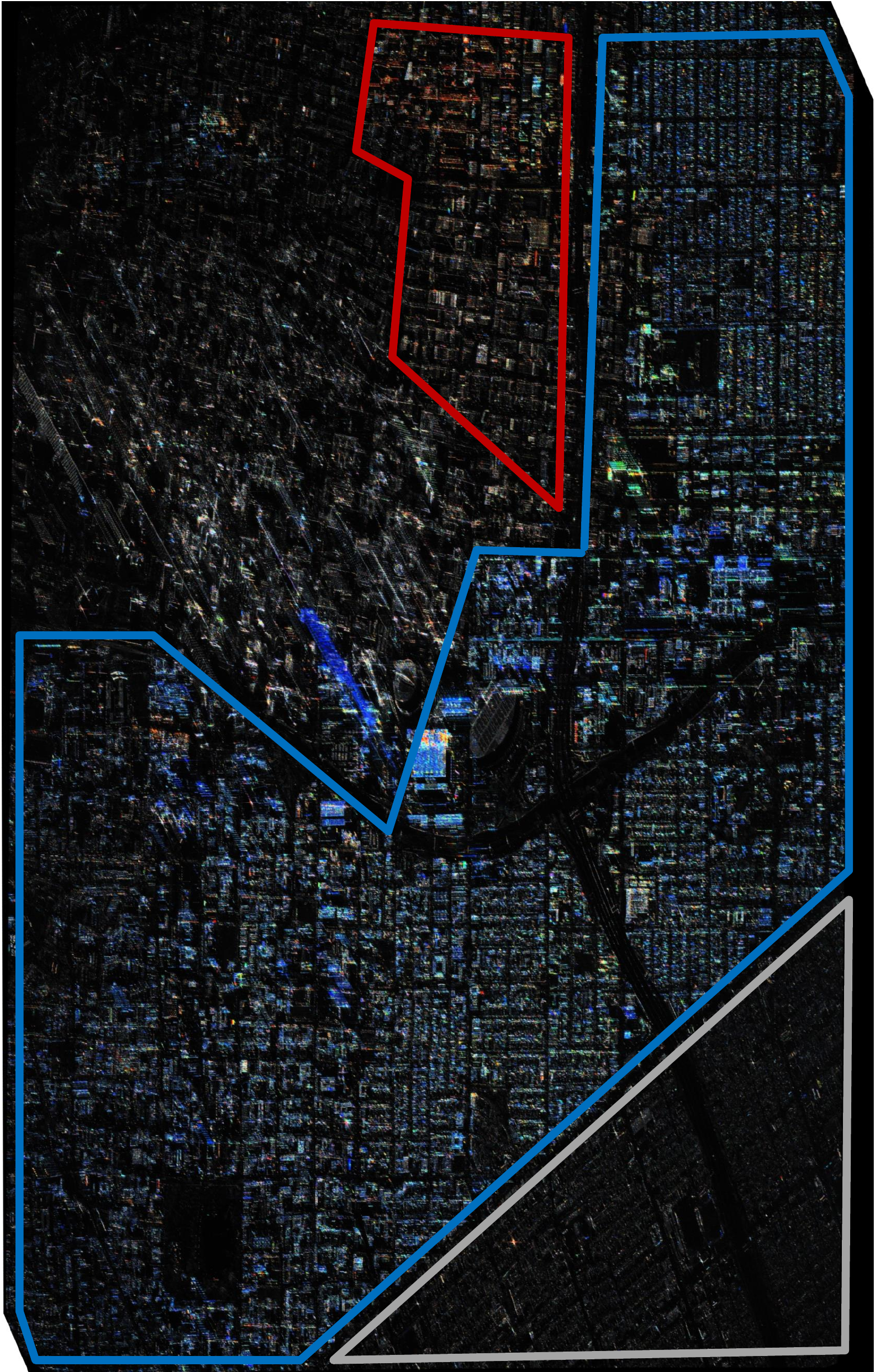}}
    \hfill
    \subfloat[Google Earth map]{\includegraphics[width=0.48\linewidth,height=10cm]{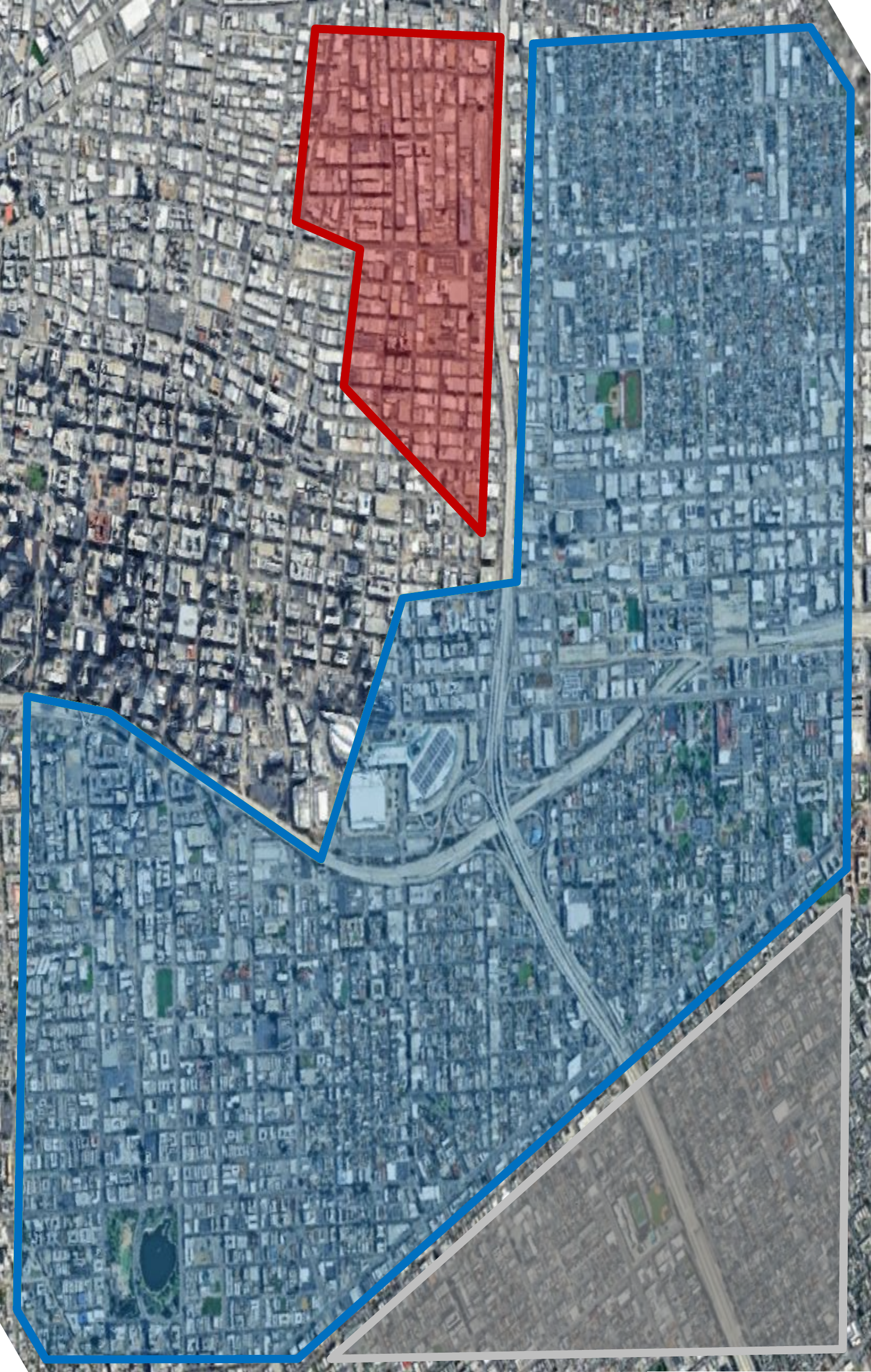}}
    \caption{\textbf{Macroscopic microwave rainbow effects in the Los Angeles urban grid.} The red and blue boxes highlight city blocks with slightly different orientations, resulting in a clear colour separation across the cityscape.}
    \label{fig:la_grid}
\end{figure*}

\begin{figure*}[t]
    \centering
    \subfloat[CSI of ocean waves, Minazuki Bay, Japan]{\includegraphics[width=0.48\linewidth,height=10cm]{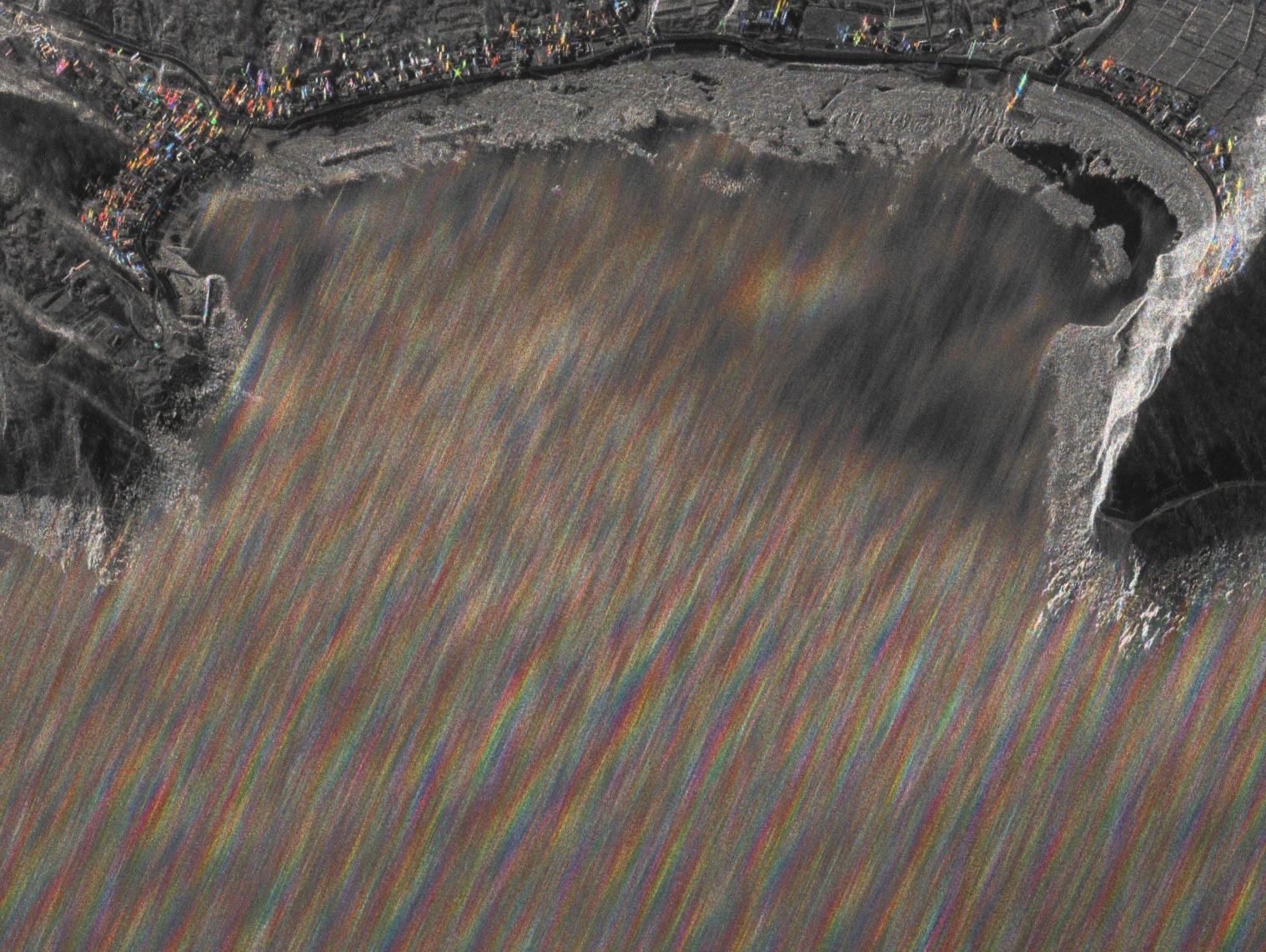}}
    \hfill
    \subfloat[Google Earth image]{\includegraphics[width=0.48\linewidth,height=10cm]{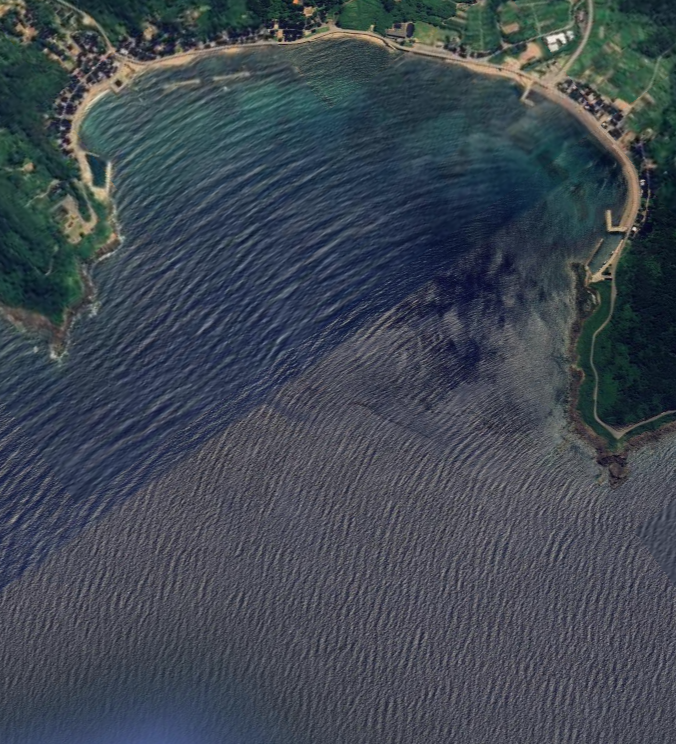}}
    \caption{\textbf{Microwave rainbow effects of ocean waves in Minazuki Bay.} The curvature of ocean wave surface have different orientation with respect to the sensor, results in RGB microwave rainbows in the CSI. Since the ocean waves is dynamic during the data acquition,  the rainbow has a smearing effect. The CSI and the Google Earth image are not captured at the same time (the sea states and wind directions
 are different), and the latter is only for illustration purpose.}
    \label{fig:noto}
\end{figure*}

\subsection*{Interpreting Microwave Rainbows in Real-World SAR Imagery}

To demonstrate the model's application to explain real-world phenomena, we now showcase the microwave rainbow across a diverse set of case studies using very-high-resolution spotlight SAR data from the Umbra open data \cite{umbra_open_data}. These examples reveal how geometry paints colours in our built environment, thereby enabling the fine-grained interpretation of infrastructure structures over large scales.

\textbf{Sliced Colours of Microwave Rainbow Unveil  Linear Structures of Building Facades, Roofs, and Bridge Constructions.} The specific orientation of a structure is the primary source of its distinct colour. Ribbed metal roof structure is a prime example, where each roof acts as a collection of linear targets. Fig.~\ref{fig:sliced_colours}a shows buildings in Las Vegas where different roof orientations produce distinct red or blue signatures, each representing a single colour from the potential rainbow spectrum. Fig.~\ref{fig:sliced_colours}b shows another example where adjacent buildings with different roof orientations produce distinct blue cyan, and yellow hues. Similarly, Fig.~\ref{fig:sliced_colours}c shows the blue signature of a building facade, a direct consequence of its specific orientation relative to the sensor's viewing angle. Fig.~\ref{fig:sliced_colours}d shows an example of steel cables of a bridge, where its three slightly curved cables with different orientation angles result in RGB colours with mild gradients.

\textbf{Zero-Order Microwave Rainbows Unveil Curvatures Structures of Road, Bridge, and Power Infrastructure.} Where  infrastructure follows a curve, our model predicts a continuous sweep of colour—a full zero-order microwave rainbow. This is consistently observed across various infrastructure types. Fig.~\ref{fig:continuous_rainbow}a shows a curved highway guardrail where the changing tangent angle maps directly to a full RGB spectrum, a real-world analog to a zero-order diffraction grating's spectral output. This effect is also ubiquitous in the sagging catenary curves of overhead power lines (Fig.~\ref{fig:continuous_rainbow}b), the curved heliostat field of a solar power plant (Fig.~\ref{fig:continuous_rainbow}c), and the stay-cables of a bridge (Fig.~\ref{fig:continuous_rainbow}d). Notably, Fig. \ref{fig:continuous_rainbow},  reveals that the heliostats exhibit a colour gradient rather than a single hue. This indicates that these heliostats are not ideal flat surfaces but possess a certain degree of curvature, indicating they are not ideal flat surfaces but possess a certain curvature. This bending phenomenon, however, is difficult to discern in optical remote sensing images—a limitation that highlights the capability of our method for fine-grained interpretation of infrastructure structures.

\textbf{High-Order Microwave Rainbows Unveil Periodic Discrete Structures of City Infrastructure.} A waterfront promenade in Dubai lined with a metallic fence produces a repeating RGB$\rightarrow$RGB colour cycle, a clear manifestation of the high-order repeating microwave rainbow predicted by our model (Fig.~\ref{fig:repeating_rainbows}a). As the fence curves, its geometric orientation changes, causing the main ($m=0$) and high-order diffraction responses ($m=\pm1, \dots$) to sweep across the Doppler spectrum. This produces the distinct, repeating diffraction pattern—an example of geometric dispersion from a discrete, periodic structure. Similarly, the seating at the Olympiastadion in Munich (Fig.~\ref{fig:repeating_rainbows}b), arranged in curved rows, behaves like a curved diffraction grating, resulting in vivid, rainbow-like patterns. 

\textbf{Macroscopic Microwave Rainbow Effects Unveil Urban Grid Structures.} On a macroscopic scale, the principles of geometric dispersion apply to the layout of entire city blocks. Fig.~\ref{fig:la_grid} shows a CSI of downtown Los Angeles where areas with different geometric orientation in their street grids exhibit clear colour separation. This large-scale observation underscores that the microwave rainbow is a phenomenon driven by the fundamental geometry of the scene, visible at all scales.

\begin{figure}[t]
  \centering
  \includegraphics[width=\columnwidth]{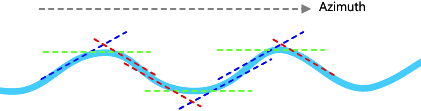}
  \caption{\textbf{Schematic of the geometric dispersion model for ocean waves.} The orientation of a wave facet relative to the sensor's azimuth direction determines its observed color. Wave surfaces oriented parallel to the flight path (green dashed lines) correspond to a central (green) Doppler response. Surfaces with positive (blue dashed lines) and negative (red dashed lines) orientations scatter energy into the edges of the Doppler spectrum, resulting in blue and red hues. The continuous change in orientation across the wave surface paints the full microwave rainbow effect.}
  \label{fig:wave_model}
\end{figure}

\textbf{Ocean Waves' Microwave Rainbows Indicate Wave Morphology and Dynamics.} Beyond static, man-made infrastructure, the principles of geometric dispersion extend to certain dynamic natural phenomena such as ocean waves. Fig.~\ref{fig:noto} shows the microwave rainbow effect on the surface of ocean waves in Minazuki Bay, Japan. Each wave acts as a series of moving, curved scatterers. This mechanism is illustrated in Fig.~\ref{fig:wave_model}, which shows how the varying local orientation of the wave surface relative to the radar's azimuth direction governs the observed color. Wave facets parallel to the flight path (green dashed lines) produce a zero-Doppler response (green), while facets with positive or negative orientations (blue and red dashed lines) scatter energy to the edges of the Doppler spectrum, resulting in blue and red hues. The constantly changing orientation of the wave facets relative to the radar's line of sight disperses the backscattered energy across the Doppler spectrum, painting the sea surface in rainbow hues. Crucially, unlike the fixed geometry of a building, the ocean surface is dynamic, evolving throughout the SAR data acquisition time. This temporal variation introduces a smearing effect to the rainbow, where the colours appear less distinct than on stationary targets, yet still clearly follow the underlying wave patterns.

\subsection*{Discussion and Summary}
This paper explains the origin of the ``microwave rainbow," a striking phenomenon where man-made structures appear systematically painted in a spectrum of colours in high-resolution radar imagery. We have demonstrated that this is not an artifact but a direct manifestation of geometric dispersion. Our findings establish that a target's own geometry effectively ``paints" its colour by acting as a physical diffraction grating, sorting reflected microwave energy by angle into the Doppler frequency spectrum.

The direct result of this physical insight is a predictive model that decodes precisely how geometry paints these colours. The framework quantitatively explains the full range of observed signatures, from the continuous colour gradients of curved targets (a zero-order diffraction effect) to the repeating spectral patterns from periodic structures (a high-order effect). Furthermore, the appearance of the microwave rainbow on dynamic ocean surfaces confirms that this geometric painting process is a universal principle, applicable to both static and moving features in our world.

This new form of microwave vision, where colour encodes geometry, suggests potential applications across multiple domains. For structural health monitoring, observing the colours painted on bridge cables could potentially reveal minute structural deformations over time. In security and intelligence, the unique colour patterns produced by periodic fences could serve as spectral fingerprints for classification. The principle may also provide a new analytical tool for planetary science, offering a way to interpret the fine-scale geometry of landforms from orbital radar data.

We acknowledge the current limitations of this approach, which in turn define directions for future work. The interpretation of colours painted in geometrically complex urban areas with significant multiple scattering requires more advanced modeling. Our model provides a forward-modeling framework to explain these phenomena, while the precise extraction of target parameters requires more refined, problem-specific modeling and model inversion. This remains a vast area for future investigation and opens up new research directions. While our work establishes geometry as the primary artist, a systematic investigation into the secondary influences of material properties and atmospheric conditions is necessary to fully calibrate this new vision.

In conclusion, our work deciphers the microwave rainbow by establishing the physical rules by which geometry paints colours in microwave vision. By providing the key to interpret these colours, we have transformed them from a visual curiosity into a quantitative measure of physical form. This establishes a new remote sensing modality and provides a fundamentally new way to perceive the geometric fabric of our world, independent of weather and daylight.

%We began with an enigma: the systematic, rainbow-like colours painting the human-made world in the latest generation of radar imagery. Here, we have resolved this puzzle by identifying geometric dispersion as the fundamental physical principle at play. Our findings establish that this phenomenon is a direct physical parallel to optical diffraction. In this 'microwave rainbow', the linear and periodic features of the built environment act as a vast diffraction grating. The very geometry of our civilization---the lines of buildings, the curves of cables, the periodicity of fences---is the engine that sorts microwave signals by angle into a spectrum of colours.
%
%The geometric-physical model we have presented provides the quantitative framework to decode these colours, transforming them from a visual curiosity into a precise metric of physical structure. This capability unlocks a new modality for remote sensing. It becomes possible to read the geometry of critical infrastructure---the subtle sag of a power line, the precise alignment of a bridge support, or the layout of an entire city grid---directly from orbit, at a planetary scale and under any weather conditions. This can be used for large-scale structural health monitoring, remote structure identification, and  planetary science. Ultimately, this work establishes the physical basis for microwave colour vision. By translating the abstract geometry of our world into a visible spectrum, we have developed a new way to see---expanding the boundaries of human vision perception.

\section*{Methods}
To explain the formation of the microwave rainbow, we must mathematically describe how a target's geometry translates its backscattered energy into a specific Doppler signature. This involved deriving the conditions for constructive interference that produce the zero-order and high-order diffraction responses, using both time-domain and frequency-domain perspectives.

\subsection*{Modelling the Squint-Angle-Dependent PSF}
The system's PSF describes the response of the SAR processor to an ideal point scatterer. For a non-zero squint angle, $\theta_{sq}$, a highly effective and physically intuitive approximation of the PSF, $h_{sq}(t_a, t_r)$, is given by:
\begin{equation}
\begin{split}
    h_{sq}(t_a, t_r) \approx\,   &\text{sinc}(t_a B_a \cos\theta_{sq}) \cdot \text{sinc}(t_r B_r \cos\theta_{sq}) \\
    &\cdot e^{j2\pi \left( (f_c \cos\theta_{sq}) t_r + f_dt_a \right)}
\end{split}
\end{equation}
where $f_{dc}$ is the Doppler centroid, $f_c$ is the radar center frequency, and $t_a, t_r, B_a, B_r$ are the time coordinates and bandwidths. The detailed derivation of this PSF is provided in the Supplementary Information. This model forms the basis for the following derivations.

\subsection*{Modelling Framework and Definitions}
We begin by defining the target and radar parameters in the imaging plane (azimuth $x$, closest slant range $y_s$). These spatial coordinates are related to the SAR system's time coordinates by $t_a = x/V$ and $t_r = 2y_s/c$, where $V$ is speed of radar and $c$ is the speed of light. A linear target in this plane is modelled as a line segment oriented at an angle $\theta_{az}$ with respect to the azimuth axis. In the time domain, this becomes:
\begin{equation}
    t_r = t_a \tan(\theta_{az}) \frac{2V}{c} = t_a K
\end{equation}
where we define the geometric constant $K = \tan(\theta_{az})(2V/c)$. The Doppler frequency, which is visualised as colour, is related to the squint angle by:
\begin{equation}
    f_{d} = \frac{2V}{\lambda} \sin(\theta_{sq})
\end{equation}

\subsection*{Derivation of Target Response in the Time Domain}
The response of a target, $g(t_a, t_r)$, is the convolution of the scene's scattering field, $s(u,v)$, with the system's PSF. We evaluate the peak response at the target's centre $(0,0)$:
\begin{equation}
    g(0,0) = \iint s(u,v) h_{sq}(-u,-v) \,du\,dv
\end{equation}
The convolving kernel $h_{sq}(-u,-v)$ becomes:
\begin{equation}
\begin{split}
    h_{sq}(-u,-v) \propto &\text{sinc}(u B_{a,eff}) \text{sinc}(v B_{r,eff}) \\
    &\cdot e^{j2\pi \left( f_{dc}u + (f_c \cos\theta_{sq})v \right)}
\end{split}
\end{equation}
where $B_{a,eff} = B_a \cos\theta_{sq}$, $B_{r,eff} = B_r \cos\theta_{sq}$, and $f_d= (2Vf_c/c)\sin\theta_{sq}$.

\textit{Continuous Target Case: The Zero-Order Response.} A continuous linear target is modelled as a line delta function: $s(u,v) = \delta(v - Ku)$. Substituting this into the response integral collapses it to:
\begin{equation}
\begin{split}
    g(0,0) \propto \int &\text{sinc}(u B_{a,eff}) \text{sinc}(Ku B_{r,eff}) \\
    &\cdot e^{j2\pi (f_d+ f_c K \cos\theta_{sq})u} \,du
\end{split}
\end{equation}
This integral's magnitude is maximised when the linear term in the exponent is zero:
\begin{equation}
    f_d+ f_c K \cos\theta_{sq} = 0
\end{equation}
This corresponds to the general diffraction condition for an integer order of $m=0$. It yields the relationship:
\begin{equation}
    \tan(\theta_{sq}) = -\tan(\theta_{az})
\end{equation}
Finally, since tan is a monotonic function, this leads to the peak Doppler response condition $\theta_{sq}= -\theta_{az}$. This fundamental relationship defines the zero-order diffraction that governs the colour of continuous targets.

\textit{Discrete Target Case: High-Order Diffraction.} A discrete linear target is modelled as a sum of $N$ equally spaced point scatterers: $s(u,v) = \sum_{n} \delta(u-u_n)\delta(v-v_n)$. The coherent sum is:
\begin{equation}
\begin{split}
    g(0,0) \propto \sum_{n=-(L-1)/2}^{(L-1)/2} e^{j2\pi(f_d+ f_c K \cos\theta_{sq})n d_u}
\end{split}
\end{equation}
This sum is maximised when the phase angle between adjacent scatterers is an integer multiple of $2\pi$, leading to the condition for constructive interference for all diffraction orders:
\begin{equation}
    (f_d+ f_c K \cos\theta_{sq}) d_u = m, \quad m \in \mathbb{Z}
\end{equation}
where $m$ is the integer diffraction order. Each value of $m$ (e.g., $m=\pm1, \pm2, \dots$) corresponds to a distinct high-order diffraction responsible for the repeating microwave rainbow, and $d_u$ is the azimuth interval of $u_n$. This leads to 
\begin{equation}\label{}
  \tan(\theta_{az}) = -\tan(\theta_{sq,m}) + \frac{m\lambda}{2d_x \cos(\theta_{sq,m})}
\end{equation}
where $d_x = V d_u$. This finally leads to Eq. (1) in the main text.

\subsection*{Derivation of Target Response in the Frequency Domain}
The same relationships can be derived from a frequency-domain perspective. The observed complex spectrum, $G(f_a, f_r)$, is:
\begin{equation}
\begin{split}
G(f_a,f_r) =\iint & s(u,v)\text{rect}\left(\frac{u}{1/B_{a,eff}}\right) \text{rect}\left(\frac{v}{1/B_{r,eff}}\right) \\
&e^{-j2\pi((f_{dc}+f_a)u + (f_c \cos\theta_{sq}+f_r)v)} \,du\,dv
\end{split}
\end{equation}

\textit{Continuous Target Case.} For a continuous target, the integral yields a sinc function whose peak response occurs when $f^{\prime}=0$ at the frequency origin, giving:
\begin{equation}
f_d+ f_c K \cos\theta_{sq} = 0
\end{equation}
This result confirms the time-domain analysis and establishes the spectral basis for the zero-order diffraction.

\textit{Discrete Target Case and Spectral Periodicity.} For a discrete target, the spectrum's magnitude is given by the Dirichlet kernel, which consists of sharp peaks when $f^{\prime} d_u = m$ for $m \in \mathbb{Z}$. Each integer $m$ corresponds to a specific diffraction order. The SAR system observes a peak frequency response when one of these spectral lines passes through the frequency origin, giving the condition for all orders:
\begin{equation}
(f_d+ f_c K \cos\theta_{sq}) d_u = m, \quad m \in \mathbb{Z}
\end{equation}
This result matches the time-domain analysis, explaining the spectral origin of both the zero-order response ($m=0$) and the high-order responses ($m \neq 0$) that form the repeating microwave rainbow. These derivations provide the complete mathematical foundation for our model.

\section*{Data availability}
The Umbra SAR data used for the case studies in this paper are part of the Umbra Open Data initiative and are publicly available at \url{https://registry.opendata.aws/umbra-open-data}. The optical images are from Google Earth. All other data supporting the findings of this study are available from the corresponding author upon reasonable request.

\section*{Code availability}
The code used for the numerical simulations and for generating the figures in this paper is available from the corresponding author upon reasonable request.

\section*{Acknowledgements}
This work was supported in part by Natural Science Foundation of China under grant No. 62301259, and the Fundamental Research Funds for the Central Universities under grant No.30924010914. The authors gratefully acknowledge Umbra Space for providing open access to their high-resolution SAR imagery.

\section*{Author contributions}
H.Y. conceived the idea, designed the study, developed the theoretical model, and performed the simulations. H.Y., and Z.H. performed data analysis. Z.L. and J.Y. supervised the research. All authors contributed to writing and reviewing the manuscript.

\section*{Competing interests}
The authors declare no competing interests.

\balance

\end{document}